\documentclass[conf]{new-aiaa}

\usepackage[utf8]{inputenc}

\usepackage{graphicx}
\usepackage{amsmath}
\usepackage[version=4]{mhchem}
\usepackage{siunitx}
\usepackage{longtable,tabularx}
\usepackage{amsthm}
\usepackage{breqn}
\usepackage{bbm} 

\usepackage{caption}
\usepackage{subcaption}
\usepackage{booktabs}
\usepackage{multirow}
\usepackage{makecell}
\usepackage{textcomp}

\newcommand{\bs}[1]{\boldsymbol{#1}} 

\newcommand{\mc}[1]{\mathcal{#1}} 

\newcommand{\ud}{\, \mathrm{d}} 

\title{Hyperspectral Lightcurve Inversion for Attitude Determination}

\author[1,*]{Simão da Graça Marto}
\author[1]{Massimiliano Vasile}
\author[2]{Andrew Campbell}
\author[2]{Paul Murray}
\author[2]{Stephen Marshall}
\author[3]{Vasili Savitski}
\affil[1]{University of Strathclyde, Mechanical and Aerospace Engineering, James Weir Building, 75 Montrose Street, Glasgow, G1 1XJ}
\affil[2]{University of Strathclyde, Electronic and Electrical Engineering, James Weir Building, 75 Montrose Street, Glasgow, G1 1XJ}
\affil[3]{Fraunhofer Centre for Applied Photonics, Level 5 Technology and Innovation Centre, 99 George Street, Glasgow, United Kingdom, G1 1RD}

\affil[*]{simao.da-graca-oliveira-marto@strath.ac.uk}

\begin{document}

\maketitle

\begin{abstract}
    Spectral lightcurves consisting of time series single-pixel spectral measurements of spacecraft are used to infer the spacecraft's attitude and rotation. Two methods are used. One based on numerical optimisation of a regularised least squares cost function, and another based on machine learning with a neural network model. The aim is to work with minimal information, thus no prior is available on the attitude nor on the inertia tensor. The theoretical and practical aspects of this task are investigated, and the methodology is tested on synthetic data. Results are shown based on synthetic data.
\end{abstract}



\section{Introduction}
\label{sec:intro}

Lightcurve inversion consists of determining information about an object in space, such as its shape, surface properties and attitude, from a single pixel observation of this object over time.
The lightcurves in classic lightcurve inversion are a mixture of light of various wavelengths into a single signal. \cite{bradley_lightcurve_nodate}
The different materials used on the surface of spacecraft have different reflectivity spectra, i.e., the way they reflect light at various wavelengths varies.
Therefore, as the spacecraft's orientation varies, different materials are visible, resulting in different spectral observations.
Previous work \cite{walker_2022_hyperspectral} made use of spectral information to infer material and component composition, and type of spacecraft information, as well as attitude.
This work builds up on the the attitude determination component of this previous work, 
by analysing time series of spectrum observations, i.e. spectral lightcurves 
to extract attitude and rotation information. 

The approaches in \cite{matsushita_light_2019, burton_two_nodate, bradley_lightcurve_nodate} infer the attitude of an object using knowledge of the dynamics, either by assuming knowledge of the inertia properties, or that the satellite is rotating about a fixed axis.
We do not include such assumptions in this work.
In \cite{burton_two_nodate}, a multiplicative Kalman filter is used, assuming fixed axis rotation.
Besides not assuming fixed axis rotation, the inertia tensor of the object is also not assumed, as this methodology is designed for scenarios where nothing is known about the attitude and rotation of the object being observed.


In this work, a purely Lambertian model is used to simulate a spectrum observation $S$, and with each surface covered in different mixtures of materials.
A convex object can be described for these purposes with a facet model, consisting of a set of $N$ faces, where the $i$-th face's normal is ${\bf n}_i$ and its average colour is ${\bf c}_i(\lambda)$. 
Average colour here means the average reflectance spectrum of the face in question multiplied by its area. For this work, the Lambertian model is used.
The intensity of the signal for wavelength $\lambda$, when the object is viewed from direction $\hat{\bf v}$ and illuminated from $\hat{\bf s}$, is then given by:
\begin{equation} \label{eq:Lambert}
    S(\lambda; \hat{\bf v}_B, \hat{\bf s}_B)
    =
    \sum_{i=1}^N
    \max(0, \hat{\bf v}_B \cdot {\bf n}_i)
    \max(0, \hat{\bf s}_B \cdot {\bf n}_i)
    {\bf c}_i(\lambda)
    ~,
\end{equation}
where certain physical quantities relating to light intensity and camera aperture have been omitted as they are not relevant for the discussion in this work.
The subscript $B$ in $\hat{\bf v}_B$ and $\hat{\bf s}_B$ indicates that these vectors are written in the body frame $B$, the same frame the normal vectors ${\bf n}_i$ are defined in. Whenever the frame these and other vectors are written in needs to be specified, this will be done with a subscript.

A consequence of the Lambertian and other models (e.g. the Lommel-Seeliger model \cite{muinonen_asteroid_2020}) is that the view and illumination angles can be swapped and this will result in the same observation, i.e. $S(\lambda; \hat{\bf v}_B, \hat{\bf s}_B) = S(\lambda; \hat{\bf s}_B, \hat{\bf v}_B)$, which results in an ambiguity in the attitude determination process. 
To this are added other ambiguities that may result from the object, even if it is asymmetric.
The bisector $\hat{\bf h}$ is the vector in between $\hat{\bf v}$ and $\hat{\bf s}$, also known as the phase vector, and defined as
\begin{equation}
    \bf \hat{h} = \frac{\hat{s} + \hat{v}}{\lVert \hat{s} + \hat{v} \rVert}
\end{equation}
Using this vector tends to be better for describing the appearance of the object to multi-spectral imaging, as suggested in \cite{nussbaum_spectral_2022}, in part due to the previously mentioned symmetry.
\section{Attitude from spectral lightcurve}
\label{sec:lightcurve}

Even for an asymmetric object, exact attitude determination is not possible for a single spectrum observation, as there are often several attitudes that result in the same observed spectrum.
These ambiguities can be partially removed by instead using lightcurve data, using the assumption that the object does not have sharp jumps in attitude or angular velocity.
Following previous work \cite{walker_2022_hyperspectral}, two methods are studied, this time applied to lightcurves instead of single spectra: Regularised least squares numerical optimisation, and machine learning.

\subsection{Regularised least squares}
\label{sec:lightcurve:reg_LS}


Works in the literature often assume constant rotation about a fixed axis (principal axis rotation) e.g. \cite{nussbaum_spectral_2022, vsilha2021light, bradley_lightcurve_nodate} but derelict satellites are likely to be tumbling (i.e. with a varying angular velocity vector).
Some works that handle these cases include \cite{matsushita_light_2019, wetterer2009}.

We also intend to assume the dynamic properties of the object, i.e. its inertia tensor $J$, are not known.
This means either combining the data by enforcing continuity in the attitude, or by explicitly including the dynamics and adding the values in $J$ as solve-fors.
In either case, this can be formulated as a problem of regularisation:
\begin{equation} \label{eq:regularization}
\begin{split}
    \min_{{\bf q}(t)} &
    \sum_i
    E(S(\hat{\bf v}_B(t), \hat{\bf s}_B(t)), \hat{S}_i)
    +
    \eta
    \int_{t_0}^{t_f}
     P({\bf q}(t)) \ud t \\
    {\rm s.t.}~& R({\bf q}(t))\hat{\bf v}_B(t) = \hat{\bf v}_I(t) \\
    ~& R({\bf q}(t))\hat{\bf s}_B(t) = \hat{\bf s}_I(t) \\
\end{split}
\end{equation}
where $\hat{S}_i$ is the observed spectrum at time $t_i$, ${\bf q}(t)$ is our attitude estimate over time in an inertial frame, and $R({\bf q})$ is the rotation operator that applies the rotation represented by ${\bf q}$. 
The position of the satellite, the Earth, and the Sun are assumed to be known precisely, such that the view and illumination vectors over time can also be written precisely in the inertial frame, $\hat{\bf v}_I(t)$ and $\hat{\bf s}_I(t)$.

The $E$ term is the least squares measurement loss function, which is the sum of squared differences between simulated and measured spectral intensity values.
The $\eta$ coefficient controls the influence of regularisation, and should be made higher the noisier the measurements are. 


The regularisation term considered in this work, $P_\alpha = \lVert\dot{\bs{\omega}}\rVert^2$, is meant to penalise changes in angular velocity.
Its time integral $J_\alpha$ is approximated as follows
\begin{equation} \label{eq:J_alpha}
    J_\alpha 
    \approx
    \frac{1}{{\Delta t}}\sum_{i=1}^{N-1}\left\lVert\bs{\omega}_{i+1}-\bs{\omega}_i\right\rVert^2
    \approx
    J_\alpha
    =
    \frac{2}{\mathrm{\Delta}t^2}{\sum_{i=1}^{N-2}\left\lVert q_{i+2}q_{i+1}^c-q_{i+1}q_i^c\right\rVert^2}
    ~.
\end{equation}
This approximation is based on the following result \cite{coutsias2004quaternions}
\begin{equation}
    \bs{\omega}_i = 
    2 \frac{d q}{d t} \Bigr|_{t=t_i} q_i^c
    \approx
    2 \frac{q_{i+1} - q_{i}}{\Delta t} q_{i}^c
    =
    \frac{2}{\Delta t} (q_{i+1}q_{i}^c - 1)
    ~,
\end{equation}
where $\bs{\omega}_i$ is given by this formula as a pure quaternion, and the time derivative of the quaternion is approximated by forward differencing.
The analytical gradient for $J_\alpha$ was also obtained, and implemented in Matlab. It is not presented here for brevity.

\subsection{Initial guess}
\label{sec:lightcurve:initial_guess}

The method previously described requires a sufficiently good initial guess in order to converge to a good solution.
Preliminary results for noiseless synthetic data with fixed axis rotation showed that if the initial guess is appropriate, the regularised LS method will converge with near exactitude to the ground truth.

A suitable initial guess is obtained using phase dispersion minimization (PDM) for the magnitude of the angular velocity, and then performing a grid search over the initial attitude and the direction of the angular velocity vector.
The de-trending for PDM is done using a high-pass filter tuned to $1/T$ where $T$ is the length in time of the tracklet.
The PDM then consists of, for every time step between 1 and $T$, performing datafolding, and measuring the variance of a sample that corresponds to the same phase angle \cite{stellingwerf1978period}.
As we have several spectra, instead of a single valued lightcurve, the data is combined by adding up these sample variances.
The local minima of these variances with respect to the period being tested are then sorted, and the 5 lowest cost candidates are then used. 

With this development, the regularised LS method often converges to the exact ground truth, but sometimes, it converges instead to a symmetric attitude history, i.e., alternative values of ${\bf q}(t)$ that result in the same spectral lightcurve observations.

\subsection{Spectral Light Curve Symmetries} 
\label{sec:lightcurve:symm}

\begin{figure}
    \centering
    \includegraphics[page=3, width=0.9\textwidth]{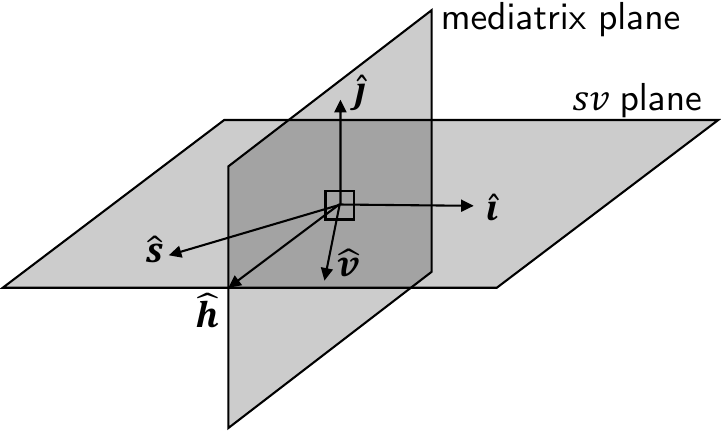}
    \caption{Diagram showing the symmetry planes and the axes of the H frame $\left(\hat{\bf h} \hat{\bf i} \hat{\bf j} \right)$}
    \label{fig:planes_HFrame}
\end{figure}

This section is concerned with the different attitude histories that result in the same light curve observations.
To study symmetries in attitude determination from lightcurves, the H frame, with the axes shown in Fig. \ref{fig:planes_HFrame}, is useful. The first axis is $\hat{\bf h}$, the third axis $\hat{\bf j}$ is along $\hat{\bf s} \times \hat{\bf v}$, and the second axis $\hat{\bf i}$ completes the right-handed triad.
In this frame, only the phase angle $\alpha$ is needed to define $\hat{\bf s}_h$ and $\hat{\bf v}_h$,
\begin{equation}
\begin{split}
    \hat{\bf s}_h &= \left[\cos{\left(\frac{\alpha}{2}\right)},\sin{\left(\frac{\alpha}{2}\right)},0\right]^T ~,\\
    \hat{\bf v}_h &= \left[\cos{\left(\frac{\alpha}{2}\right)},-\sin{\left(\frac{\alpha}{2}\right)},0\right]^T
    ~.
\end{split}
\end{equation}
Therefore, the vectors $\hat{\bf s}_B$ and $\hat{\bf v}_B$ can be recovered from $\alpha$ and $R_{BH}$, the rotation that transforms from the H to the B reference frames \footnote{Assuming $\hat{\bf s}\cdot\hat{\bf v}\neq \pm 1$}.
The spectrum function $S$, before written as $S = S(\lambda; \hat{\bf v}_B, \hat{\bf s}_B)$, can now also be written as a function of $\alpha$ and $R_{BH}$, i.e. $S = S(\lambda; R_{BH}, \alpha)$.



As shown before, any transformation that either swaps the $\hat{\bf s}_B$ and $\hat{\bf v}_B$ vectors or keeps them in place results in the same spectrum.
There are three orthogonal transformations $T_{H_i} \in \mc{T}_H$ that achieve this (four, if the identity transformation, $I$, is included).
Two are reflections about the planes in Fig \ref{fig:planes_HFrame}, $T_{H_1}$, about the mediatrix plane, and $T_{H_2}$, about the $sv$ plane. The other is a 180$^\circ$ rotation about $\hat{\bf h}$, $T_{H_3}$, which is also the composition of those two reflections, $T_{H_3} = T_{H_1}T_{H_2}$. 
Note that these symmetries are completely independent of the target object being observed and are purely a consequence of the scattering law being used.

The symmetry group $\mc{T}_H$ can now be defined, regardless of the object in question, as $\mc{T}_H = \{I,~ T_{H_1},~ T_{H_2},~ T_{H_3}\}$.

The body may also have symmetries $T_B \in \mc{T}_B$, such that, coming back to the previous formulation of $S$, we may write $S(\lambda; \hat{\bf v}_B, \hat{\bf s}_B) = S(\lambda; T_B \hat{\bf v}_B, T_B \hat{\bf s}_B)$.
Note that these are symmetries of the facet model, which may or may not be symmetries of the actual object.
Written in compact form 
$S(\lambda; R_{BH}, \alpha) = S(\lambda; R_{BH} T_B, \alpha)$.
Combining the H-frame symmetries with the body symmetries, 
\begin{equation}
    S = S(R_{BH}, \alpha)
    = S(T_H R_{BH} T_B, \alpha) ~\forall T_H \in \mc{T}_H~,~\forall T_B \in \mc{T}_B
\end{equation}
where, as is done in group theory, the symmetry groups $\mc{T}_H$ and $\mc{T}_B$ always contain at least the identity transformation $I$.
The spectrum $S$ here was parameterised with $R_{BH}$, which is to be left-multiplied with (column) vectors on the $B$ frame or right-multiplied with (row) vectors on the $H$ frame, which is why $R_{BH}$ right-multiplies $T_H$ and left-multiplies $T_B$.
It does not matter whether, for a particular symmetry $T_H$ or $T_B$, it or its transpose should be used in the formula above, since if $\mc{T}$ is a symmetry group, $T\in\mc{T}$ iff $T^{-1}\in\mc{T}$, a condition which is one of the requirements for $\mc{T}$ to be called a ``group'' in the abstract algebra sense.

The rotation $R_{BH}$ represents the attitude of the body relative to the H frame. 
The transformation $R_{IH}$ is obtained from the vectors $\hat{\bf v}_I$ and $\hat{\bf s}_I$, which are known since the geometry of the observation is known.
This allows recovering the object's attitude in the inertial frame $R_{BI}$ as
\begin{equation}
    R_{BI} = R_{IH}^T R_{BH}
\end{equation}
remembering that for orthogonal transformations $R$, it holds that $R^T = R^{-1}$, so in our notation $R_{IH}^T = R_{HI}$.

Therefore, for any real attitude history $R_{BI}(t)$, the attitude histories given by the following set are indistinguishable in terms of the generated spectral lightcurve
\begin{equation} \label{eq:symm_group}
    \mc{R}(R_{BI}) = 
    \left\{
    \begin{array}{cc}
    R_{BI}^S = R_{IH}^T T_H R_{IH} R_{BI} T_B
    :
    \det(R_{BI}^S) = 1
    ,~\vspace{0.7em}\\
    \forall T_H \in \mc{T}_H
    ~,~
    \forall T_B \in \mc{T}_B
    \end{array}
    \right\}
    ~.
\end{equation}
The condition $\det(R_{BI}^S) = 1$ is to enforce $R_{BI}^S$ to be a rotation (i.e., an element of $SO(3)$), which is necessary for it to describe the attitude of the object.
This condition is equivalent to $\det(T_H)\det(T_B)=1$, in other words, either both $T_H$ and $T_B$ are rotations (with $\det = 1$) or they are both reflections (with $\det = -1$).
Of the symmetries in $\mc{T}_H$, only $T_{H_3}$ is a rotation, i.e., has determinant equal to one.
This has the result that if the object is completely asymmetric, i.e., $\mc{T}_B$ contains only identity, $\mc{T}_B = \{I\}$, then 

\begin{equation}
    \mc{R}(R_{BI})
    =
    \left\{
    R_{BI} ~,~ R_{IH}^T T_{H_3} R_{IH} R_{BI}
    \right\}
\end{equation}

If, on the other hand, $\mc{T}_B$ contains a symmetry $T_{BR}$ which is a reflection, i.e. $\det{T_{BR}} = -1$ and $\mc{T}_B=\{I,~ T_{BR}\}$, then the set $\mc{R}(R_{BI})$ now has four elements, two of them combining $T_{BR}$ with the reflections in $\mc{T}_H$ which are $T_{H_1}$ and $T_{H_2}$. In that case,

\begin{equation}
    \mc{R}(R_{BI})
    =
    \{
    R_{BI} ~,~ R_{IH}^T T_{H_3} R_{IH} R_{BI} 
    ~,~
    R_{IH}^T T_{H_2} R_{IH} R_{BI} T_{BR}
    ~,~
    R_{IH}^T T_{H_1} R_{IH} R_{BI} T_{BR}
    \}
    ~.
\end{equation}

The previously mentioned cases where lightcurve inversion resulted in an attitude history that did not match the ground truth were found to match an attitude history given by the above expression.


It is also interesting to see what the above means for the angular velocity vector.
There are two ways to define it - in the body frame or in the inertial frame.\footnote{Note this choice does not affect the value of the regularization term $J_\alpha$, in Section \ref{sec:lightcurve:reg_LS}.}
The angular velocity of the object in the inertial frame is obtained from the following relation
\begin{equation} \label{eq:omega_def}
    [\bs{\omega}_{BI}\times]
    =
    \dot{R}_{BI} R_{BI}^T
    ~,
\end{equation}
where $[\bs{\omega}_{BI}\times]$ is the linear operator (i.e. matrix) such that $[\bs{\omega}_{BI}\times] {\bf x}$ is equal to the cross product $\bs{\omega}_{BI}\times {\bf x}$, 
and $\dot{R}$ denotes the element-wise time derivative of $R$.
Vectors $\bs{\omega}_{IH}$ and $\bs{\omega}_{BH}$ are defined in the same way.
Defining $\bs{\omega}_{IB}$ in the same way corresponds to the negative of the body's angular velocity in the body frame, i.e.
\begin{equation}
    \bs{\omega}_{IB}
    =
    - R_{BI}^T \bs{\omega}_{BI}
    ~.
\end{equation}

So far, the time dependencies of the various rotations were omitted.
The set $\mc{T}_H$ only contains 4 distinct elements, so the fact that $R_{BI}(t)$ must be continuous means  $T_H$ is constant.
If the set $\mc{T}_B$ is finite, $T_B$ must also be constant if one assumes that the elements of $\mc{R}$ are continuous functions of time. 
Otherwise, e.g. if the object has cylindrical symmetry, $T_B$ can be time-varying, and the process of attitude determination becomes significantly more challenging. This possibility is not considered in this work at the moment.

Including the time dependency, the elements of set $\mc{R}(R_{BI})$ are written as
\begin{equation}
    R_{BI}^S(t) = R_{IH}(t)^T T_H R_{IH}(t) R_{BI}(t) T_B
    ~,
\end{equation}
which by application of Eq. (\ref{eq:omega_def}) results in, 
\begin{equation}
    \bs{\omega}^S_{BI} (t)
    =
    \bs{\omega}_{HI} (t)
    +
    R_{IH}^T T_H R_{IH} 
    \left(\bs{\omega}_{BI} (t) - \bs{\omega}_{HI} (t)\right)
    ~,
\end{equation}
and the set $\mc{W}(R_{BI})$ is defined similarly to $\mc{R}(R_{BI})$, but containing $\bs{\omega}^S_{BI}$, i.e.:
\begin{equation} \label{eq:symm_group_wBI}
    \mc{W}(\bs{\omega}_{BI}(t)) = 
    \left\{
    \begin{array}{cc}
    \bs{\omega}_{HI}(t) 
    +
    R_{IH}^T(t) T_H R_{IH}(t) 
    \left(\bs{\omega}_{BI}(t)  - \bs{\omega}_{HI} (t)\right)
    ,~\vspace{0.7em}\\
    \forall T_H \in \mc{T}_H : \exists T_B \in \mc{T}_B : det(T_H T_B) = 1
    \end{array}
    \right\}
    ~.
\end{equation}

Even if $\bs{\omega}_{BI}$ is constant, $\bs{\omega}^S_{BI}$ may not be because $R_{IH}$ and $\bs{\omega}_{HI}$ are not constant. 
Therefore, this ambiguity could be removed if one could assume fixed axis rotation, or knowledge of the attitude dynamics of the object, which is not the case for this work. This is also why in other works the authors are able to obtain the exact attitude despite these symmetries.
Note that $T_B$ does not appear in the expression for $\bs{\omega}^S_{BI}$. (It would appear in a similar expression for $\bs{\omega}^S_{IB}$ however.)
This means $\mc{W}(R_{BI})$ contains at most four elements, regardless of the symmetries of the body (if these are not continuous).
However, the presence of symmetries $T_B$ still affects the number of elements of $\mc{W}(R_{BI})$, because if any of them has $\det(T_B)=-1$, then all four symmetries $\mc{T}_H$ can be used, due to the condition $\det(R_{BI}^S)=1$.
This phenomenon was also confirmed with synthetic results.

It is useful to write $\bs{\omega}$ in the H frame, 
\begin{equation}
    \bs{\omega}_{BH}
    =
    R_{IH}\left(\bs{\omega}_{BI} - \bs{\omega}_{HI} \right)
    ~.
\end{equation}
Writing the symmetric versions corresponding to the elements of $\mc{R}(R_{BI})$, 
\begin{equation}
\label{eq:wBH}
    \bs{\omega}_{BH}^S
    =
    T_H R_{IH}\left(\bs{\omega}_{BI} - \bs{\omega}_{HI} \right)
    ~.
\end{equation}
Because the transformations $T_H$ can only change the sign of the $y$ and $z$ components of $\bs{\omega}_{BH}^S$, without altering its values, a ML model should be able to predict the absolute values of these  $y$ and $z$ components, even if not the sign. Therefore, the model can be trained with vector $\bs{\omega}_{BH}$ with those components replaced by their absolute values to avoid issues caused by the symmetries. This is tested in Section \ref{sec:results:ML:symm}.
Throughout the derivations above, well-known identities were used, all of which are commonly found in the literature, such as \cite{coutsias2004quaternions}.


\subsection{Solution Search with Differential Dynamic Programming}
\label{sec:lightcurve:DDP}

Differential dynamic programming (DDP) is a strategy for efficiently optimising trajectories with respect to some control law. \cite{mayne1973DDP, tassa2012ddp}
The trajectory is discretised as a sequence of states ${\bf x}_i$ and control variables ${\bf u}_i$ for time steps ${\bf t}_i$ which correspond to spectral observations $\hat{S}_i$, and the following loss function is optimised recursively
\begin{equation} 
\label{eq:DDP_loss}
    V
    =
    \sum_{i} l_k({\bf x}_i, {\bf u}_i)
    ~,~
    {\bf x}_{i+1} = F({\bf x}_i)
\end{equation}
where $l_i$ is a path loss function and $F$ is the state transition function.

The Matlab implementation of DDP used in this work can be found in \cite{yuval2023ddp}.

The loss function in Eq. (\ref{eq:regularization}) can be adapted to the form of Eq. (\ref{eq:DDP_loss}) by writing ${\bf x}_i = \begin{bmatrix}
        {\bf q}_i \\ \bs{\omega}_i
    \end{bmatrix}$
and
\begin{equation}
    l_i({\bf x}_i, {\bf u}_i)
    =
    l_E({\bf q}_i) + l_\alpha({\bf u}_i)
    =
    \lVert
    \hat{S}_i
    -
    S(R({\bf q}_i)^T\hat{\bf v}_I(t_i), R({\bf q}_i)^T\hat{\bf s}_I(t_i))
    \rVert^2
    +
    \eta_{\alpha}
    \lVert
    {\bf u}_i
    \rVert^2
\end{equation}

Here, $\bf u$ is the angular acceleration, $l_E$ is the sum of squared differences measurement loss and $l_\alpha$ is the regularisation loss, which approximates $P_\alpha$.
In this problem, there is no loss on the final state as there often is in DDP.
The state transition function is
\begin{equation}
    {\bf q}_{i+1} = {\bf q}(\bs{\omega}_i){\bf q}_{i}
    ~,~
    \bs{\omega}_{i+1} = 
    \bs{\omega}_i + (t_{i+1}-t_i) {\bf u}_i
\end{equation}
where ${\bf q}(\bs{\omega}_i)$ is the quaternion with real part $\cos(\beta/2)$ and vector part $\sin(\beta/2) \bs{\omega}_i / \lVert \bs{\omega}_i \rVert$ with $\beta = \arccos((t_{i+1}-t_i)\lVert \bs{\omega}_i \rVert)$, i.e. it describes the rotation with angular velocity $\bs{\omega}_i$ after $(t_{i+1}-t_i)$ time passes, absent any acceleration.
This state transition function is an easy to compute approximation.

\subsubsection{Best first search}

Given the significant speed-up achieved by DDP, a more thorough and less informed search for an initial starting position for optimisation can be pursued.
A strategy was devised that does not make use of knowledge of the period of rotation from PDM, and does not form the initial guess based on an assumption of fixed axis rotation.

A set of initial orientations ${\bf q}_0$ is obtained by uniformly sampling the set of quaternions, using the same strategy employed in \cite{ricciardi_improved_2019} to obtain a set of well spaced search directions.
A graph search approach, named best first search \cite{pearl1984heuristics}, is used to obtain a solution.
Using terminology associated with these methods, consider a node to be a trajectory defined by the states ${\bf x}_k$ and control variables ${\bf u}_k$ for $k = {1,\ldots,n}$.
Expanding such a node consists of using this trajectory as an initial guess to optimise (with DDP) the trajectory defined for $k = {1,\ldots,n,n+1}$.
The initial guess for ${\bf x}_{n+1}$ is ${\bf x}_n$, and for ${\bf u}_k$ it is zero.
The initial nodes are formed by combining the ${\bf q}_0$ with a zero angular velocity.
Repeatedly expanding one of the initial nodes $n$ times results in the full trajectory we seek.
At each expansion, the states and controls for previous time steps may change. 
It is not correct to see the states and controls at each time step as the nodes, but rather the whole trajectory segments.

In best first search \cite{pearl1984heuristics} one starts with a list of nodes and expands the one with the lowest cost until the full path is obtained.
To obtain an initial estimate of the cost, we start by expanding all the initial nodes ${\bf q}_0$ into trajectories with $n=2$ and store them in a priority queue.
Thenceforth the algorithm proceeds as a best first search algorithm does until a final trajectory is obtained.
Note that the total number of nodes never exceeds the initial number, since each ``expansion'' produces one node from a previous node. This could be changed in the future by using a sample of initial guesses for ${\bf x}_{n+1}$ and ${\bf u}_{n+1}$ instead of the values described above.

An issue with this approach is that once a promising node has been expanded to a high value of $n$, nodes that have not yet been expanded by much in comparison will be selected for expansion, which tends to be fruitless since the cost increases to be higher than the best case before the high value of $n$ is reached.
This is avoided in best-first search methods \cite{pearl1984heuristics} by adding a heuristic.
The theory behind these methods informs that this heuristic should be "admissible", i.e. a conservative estimate of the cost to be incurred if this node is expanded until the end of the trajectory. 
To that end, the following heuristic is proposed
\begin{equation}
    h({\bf x}_1, {\bf u}_1, \ldots, {\bf x}_n, {\bf u}_n)
    =
    \frac{V}{a n}
    \max(0, n_{\rm max} - n - n_0) 
    \rho^{\max(0, k - k_0)}
\end{equation}
where $V$ is the node's total cost, $n_{\rm max}$ is the length of the most expanded node, $a=4$ is a scaling factor to help ensure the estimate is conservative, $n_0$ is the minimum difference for this heuristic to be non-zero.
The remaining terms form an ``impatience'' cost, intended to penalise more strongly the evaluation of low $n$ nodes when the algorithm has been running for a long time, measured by the number of iterations performed $k$. This factor is in use when $k\geq k_0 = 1000$, and $\rho = 1.001$.

Many assumptions relied on to prove best-first search's optimality are not valid here. For example, expanding a node may decrease its cost if the previous optimisation converged to a suboptimal trajectory, different paths may end up converging to the same trajectory, and there is no guarantee the heuristic here proposed will be a conservative estimate of the actual cost.

\subsection{Attitude Estimation with Machine Learning}
\label{sec:lightcurve:ML}

The machine learning strategy consists of a densely connected neural network trained using Adam optimisation and dropout neurons, following the work in \cite{walker_2022_hyperspectral}. 
All machine learning models were implemented and trained using the Keras module in Python.
The input to the model may be the time series of spectral observations $\hat{S}_k$ and vectors  $\hat{\bf s}(t_k)$, $\hat{\bf v}(t_k)$, all combined into a single input vector fed into the neural network.
If there are 4 spectral components and 200 time steps, this results in $(4+6)\times200=2000$ input parameters.
The output is the angular velocity vector at the middle of the trajectory.

There is one issue with this approach. If $T$ is an orthogonal transformation, replacing $\hat{\bf s}_I(t)$, $\hat{\bf v}_I(t)$, by their transformations $T \hat{\bf s}_I(t)$, $T \hat{\bf v}_I(t)$, for the same observed spectal lightcurve $\hat{S}_i$, will correspond to the same $\bs{\omega}_{IB}$ or to a similarly transformed $T \bs{\omega}_{BI}$.
Furthermore, during a tracklet, the vector $\hat{\bf s}_I$ varies very slightly, since the length of a tracklet is typically measured in minutes, while $\hat{\bf s}_I$ takes a full year to complete one revolution.
Reducing variation in the data can improve the efficiency of the machine learning process, and in this case it is also possible to reduce the number of inputs to the model.

\begin{figure}
    \centering
    \includegraphics[width=\textwidth]{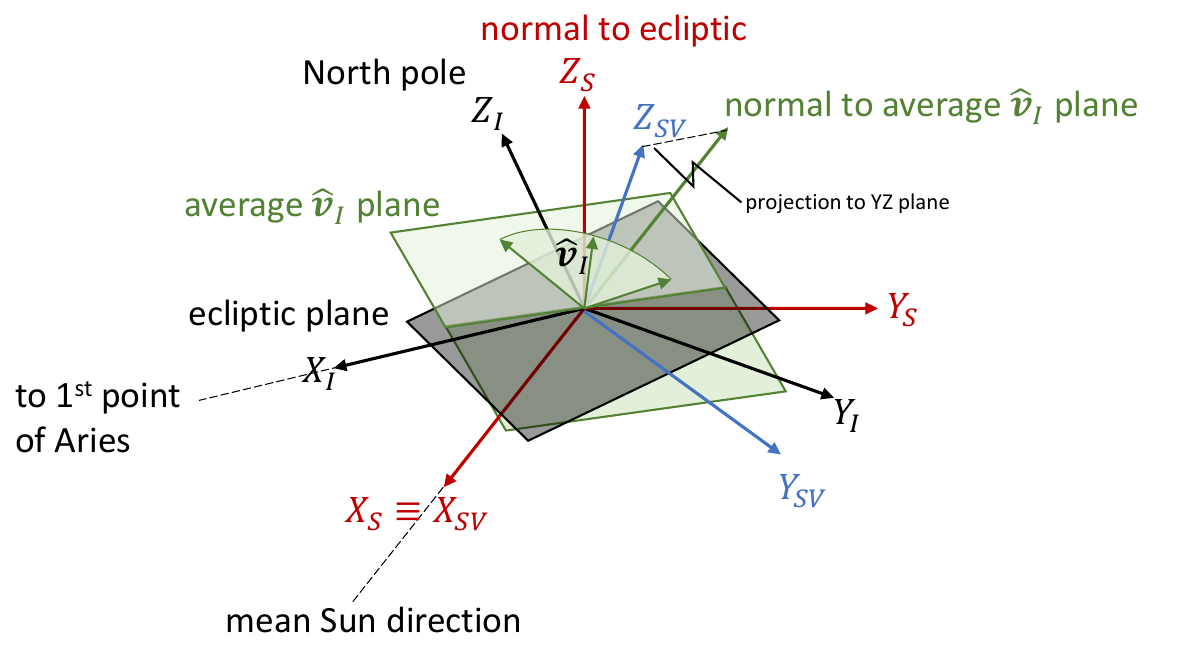}
    \caption{Diagram illustrating the various inertial reference frames}
    \label{fig:XYZ}
\end{figure}

\begin{figure}
    \centering
    \includegraphics[width=0.8\textwidth]{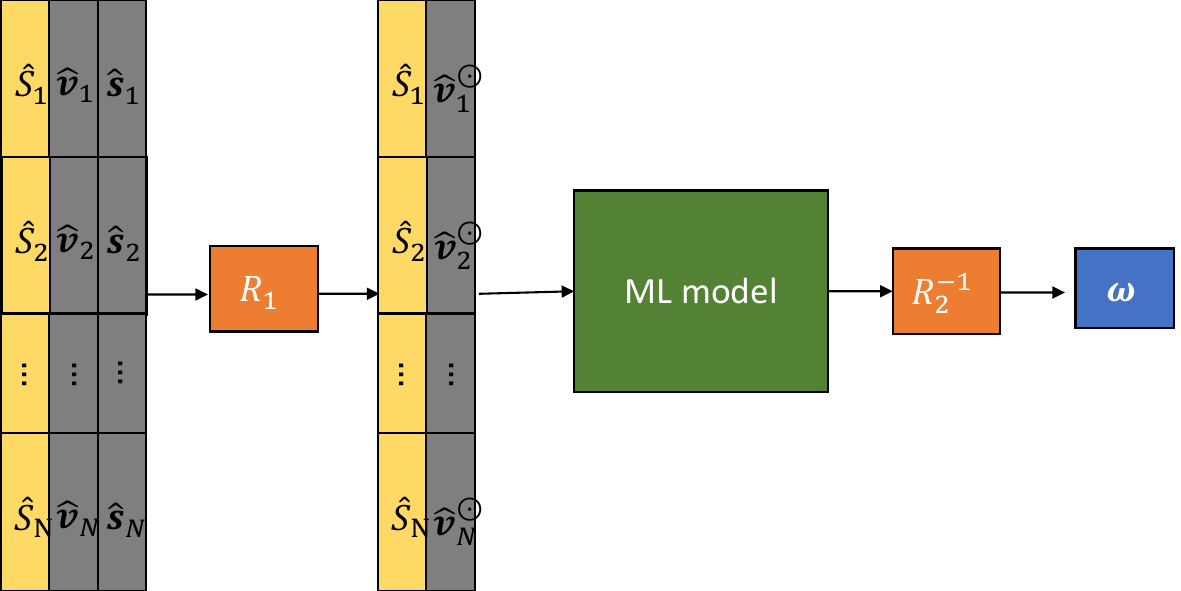}
    \caption{Diagram illustrating the overall ML approach.}
    \label{fig:ML_diagram}
\end{figure}

This is done by choosing a reference frame, which has the $x$ axis, $X_S$, aligned with the mean sun direction over the length of the tracklet. This is not enough to define a reference frame.
Two reference frames are used, the ``S'' and ``SV'' frames. 
Both of these reference frames are inertial frames. In this discussion, the term ``inertial (reference) frame" refers to the Earth centered inertial (ECI) frame, with its $x$ axis pointing towards the first point of Aries, unless noted otherwise. 
These frames are illustrated in Fig. \ref{fig:XYZ}, where $X_IY_IZ_I$ are the axes of the ECI frame, $X_SY_SZ_S$ corresponds to the S frame, and $X_{SV}Y_{SV}Z_{SV}$ to the SV frame.
In the first approach, the $Z_S$ axis corresponds to the normal to the ecliptic plane.
The reference frame thus defined is termed S.
In the second, the $Z_{SV}$ axis is the projection onto the $Y_IZ_I$ plane of ECI of the normal to the mean plane that contains $\hat{\bf v}$. This is done to further constrain the variation of vector $\hat{\bf v}$ to be mostly in the x and y directions. This frame is termed SV.
This approach is illustrated in Fig. \ref{fig:ML_diagram}.

In both frames S and SV, only the vector $\hat{\bf v}$ needs to be input to the model, since $\hat{\bf s} \approx [1,0,0]^T$,
and the $x$ component of both $\hat{\bf v}_S$ and $\hat{\bf v}_{SV}$ is the cosine of $\alpha$.
If the model is predicting the attitude $R_{BI}$ or the angular velocity in the inertial frame $\bs{\omega}_{BI}$, these also need to be updated similarly.
If instead $\bs{\omega}_{IB}$ is being predicted, this vector does not need to be transformed.
The ML models in the tests conducted so far only predict the angular velocity vector at the middle of the observation tracklet.
Which reference frame to use, whether the change of reference frame mentioned is useful, and which architecture to use are decided by tests, shown in Section \ref{sec:results:ML}.

Another issue which may affect the quality of the model are the symmetries, discussed in Section \ref{sec:lightcurve:symm}.
Since the symmetries only affect the signs of the $y$ and $z$ components of $\bs{\omega}_{BH}$, the vector $\bs{\omega}_{BH}^S$ is defined as $\bs{\omega}_{BH}$ with its $y$ and $z$ components replaced by their absolute values.
Some tests were also performed where the model was trained on the absolute vales of all components, written as $\lvert \bs{\omega}_{BI} \rvert$ and $\lvert \bs{\omega}_{BH}\rvert$ (These are not the norms of the vectors, which are written as $\lVert \bs{\omega}_{BH}\rVert$, but the vectors containing the absolute values of each element).



\subsubsection{Network Architecture}

The models under consideration are all densely connected neural networks.
Several models are defined, termed $M_0$ through $M_4$.
Let $W(n)$ designate a fully connected hidden layer with $n$ neurons, and $D(p)$ a dropout layer \cite{srivastava2014dropout} with dropout probability $p$.
All fully connected layers have ReLU activation layers except for the final layers which do not have any activation function.
The input of the model, $X$, contains $\hat{S}_i$, $\hat{\bf v}(t_i)$ in either the ECI, S, or SV frames, and in the case of ECI, $\hat{\bf s}_I(t_i)$ is also used. The output $Y$ is the angular velocity vector $\bs{\omega}$ at the midpoint of the tracklet in the ECI or H frames, with some or all of its elements replaced by their absolute values as discussed previously.
The models are then written as
\begin{align}
    &M_0 : X \rightarrow W(600) \rightarrow Y \\
    &M_1 : X \rightarrow W(600) \rightarrow W(50) \rightarrow Y \\
    &M_2 : X \rightarrow W(600) \rightarrow D(0.2) \rightarrow W(50) \rightarrow Y \\
    &M_3 : X \rightarrow W(1000) \rightarrow D(0.2) \rightarrow W(100) \rightarrow D(0.2) \rightarrow Y \\
    &M_4 : X \rightarrow W(1000) \rightarrow D(0.2) \rightarrow W(200) \rightarrow D(0.2) \rightarrow W(50) \rightarrow D(0.2) \rightarrow Y 
\end{align}


The training is done using the Adam optimiser \cite{kingma2014adam}, with the mean squared error (MSE) loss function.
The stopping conditions for the training were a maximum number of epochs of 10000 in addition to stopping if the validation loss does not decrease by more than $10^{-4}$, in relative terms, for over 100 epochs.

\section{Results}
\label{sec:results}

The methods described previously were tested, and the results obtained are presented in this section.
But first, two datasets are described.

\subsection{Dataset 1}

A cube object was considered, with each face composed of a different material, and assuming a perfect identification of these materials. The resulting observations can be simulated using the model in Eq. (\ref{eq:Lambert}).
This facet model consists of 6 vectors, consisting of the three orthogonal axes of the body frame plus their negatives.
Each facet contains a different material, such that ${\bf c}_k$ is a one-hot vector, i.e. it has a one in the $k$-th element and zeroes elsewhere.
Therefore, this object is asymmetric, which means $\boldsymbol{\omega}_{IB}$ is theoretically unique.
From the discussion in Section \ref{sec:lightcurve:symm}, however, the same is not the case for $\boldsymbol{\omega}_{BI}$.

The observation tracklets $\hat{S}_i$ were then obtained as follows.
The satellite's orbit was simulated with the Keplerian elements of STARLINK-1272 as of 2023/03/12, using TLE data from n2yo.com, and these elements were propagated in a purely Keplerian way, i.e., ignoring perturbations.
Thus, the satellite's position relative to a ground station in Glasgow was calculated for the 8 year period encompassing 2023-2030.
All observation windows are then found for this time period. An observation window here means the period of time during which observation of the satellite is possible. This is based on three conditions:
\begin{itemize}
    \item The satellite is 10 degrees above the horizon as seen from the ground station;
    \item The satellite is out of the Earth's shadow;
    \item The ground station is at night (i.e., Sun is below horizon. Twilight conditions not considered).
\end{itemize}

Afterwards, the attitude history is simulated. The satellite is assumed to have a spherically symmetric inertia tensor, so the rotation axis is fixed.
The angular velocity was chosen as 2 rotations per minute (RPM).
For each observation window, $N_0 = 100$ different initial orientations and axis of rotation are selected randomly, and the resulting spectra are simulated.

The values of the observed spectrum $\hat{S}_i$, the observation vectors $\hat{\bf v}_I(t_i)$ and $\hat{\bf s}_I(t_i)$, the ground truth attitude history in the form of quaternions ${\bf q}_{GT}(t_i)$, and the angular velocity vectors $\boldsymbol{\omega}_{BI}$ and $\boldsymbol{\omega}_{IB}$, are stored for every tracklet. 
In the following discussion, we refer to a tracklet as $\hat{S}_i$, but all attitude determination methods always have access to the vectors $\hat{\bf v}_I(t_i)$ and $\hat{\bf s}_I(t_i)$, or a transformed version thereof.

In addition to this dataset, we also define dataset 1A as being obtained the same way, but adding symmetry by making the materials of two opposing faces the same. 
This symmetry is represented by transformation $T_{BR}$, which corresponds to a reflection about the $xz$ plane. 

\subsection{Dataset 2}

A more complex dataset was also obtained, to more thoroughly test the lightcurve inversion methods.
It was obtained following the same methodology as dataset 1, apart for what is described in this subsection.

\begin{figure}
    \centering
    \includegraphics[width=0.5\textwidth]{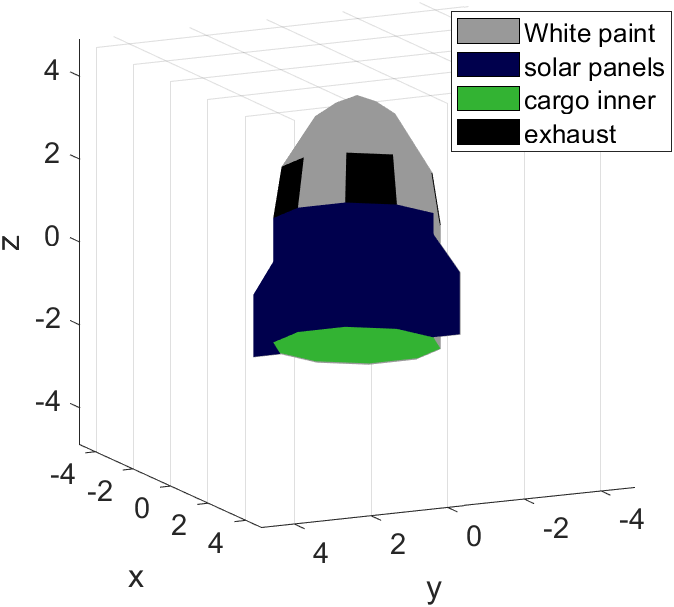}
    \caption{Model used to simulate the spectral observations for dataset 2.}
    \label{fig:dragon_model}
\end{figure}

The simulated object is a model of the Dragon capsule in Fig. \ref{fig:dragon_model}, with 4 different materials.
It also has approximately a symmetry about the $xz$ plane, represented with the same matrix $T_{BR}$ as the previous dataset. A reflection about this plane does not result in the exact same object due only to discretisation of the curved surfaces of the object.
Its orbital elements are chosen randomly from a list of epheremids for LEO objects\footnote{Sourced from www.space-track.org}.

The rotation axis is not fixed. Instead, for this dataset, torque-free motion was assumed, with three different options for the inertia tensor $J$. These are
\begin{equation}
    J^{(1)} = \begin{bmatrix}
        1&0&0\\0&1&0\\0&0&1
    \end{bmatrix}
    ~,~
    J^{(2)} = R \begin{bmatrix}
        \gamma&0&0\\0&1&0\\0&0&1
    \end{bmatrix}
    R^T
    ~,~
    J^{(2)} = R \begin{bmatrix}
        \gamma&0&0\\0&\gamma&0\\0&0&1
    \end{bmatrix}
    R^T
    ~,
\end{equation}
where $R$ is a randomly obtained rotation matrix and $\gamma = 1.1$.
For torque-free dynamics, scaled variants of the inertia tensor result in the same attitude trajectory, which is why the units are not specified.
%
%
%
The magnitude of the angular velocity vector is chosen randomly for each tracklet, using the following formula
\begin{equation}
    \log \left\lVert \omega_{BI, 0} \right\rVert
    =
    u
    {(\log \omega_{\rm max} - \log \omega_{\rm min}) + \log \omega_{\rm min}},
\end{equation}
where $u$ is randomly selected from a uniform distribution between 0 and 1, $\omega_{\rm min} = 0.5$ RPM and $\omega_{\rm max} = 2$ RPM. 
In other words, the angular velocity ranges between 0.5 and 2 rotations per minute (RPM).

\subsection{Illustration of Symmetries in Light Curve Inversion}

As mentioned, dataset 1A is obtained for a cube object with all faces different from each other except for two opposite faces which are the same. This means the set $\mc{T}_B$ contains identity as well as a planar reflection $T_{BR}$.



\begin{figure}[thb]
    \centering
    \includegraphics[width=0.8\textwidth]{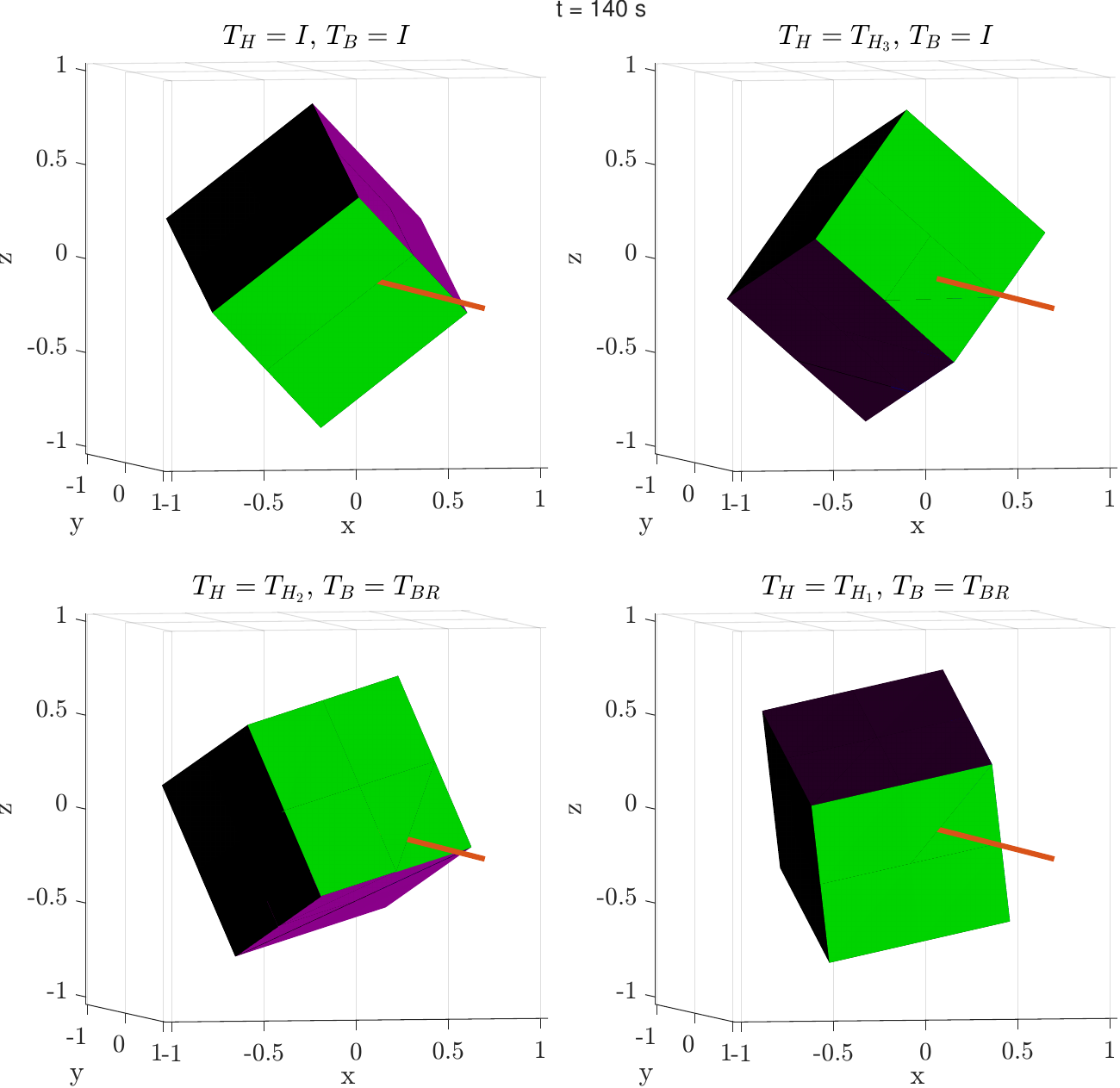}
    \caption{Appearance of cube object for attitudes in $\mc{R}$ at $t=140s$. Faces colours are illustrative only, and do not correspond to the spectral reflectance of these faces, but the brightness of the faces is accurate to the viewing conditions}
    \label{fig:symm_2}
\end{figure}

Figure \ref{fig:symm_2} shows, at some instant, the object as it would be seen from an observer at direction $\hat{\bf v}$, illuminated from $\hat{\bf s}$, which is marked as an orange line, for different attitudes that result in the same spectrum observation. 
These are the attitudes contained in set $\mc{R}$ defined in Eq. (\ref{eq:symm_group}). They are identified in the figures by the different elements $T_H$ and $T_B$, of $\mc{T}_H$ and $\mc{T}_B$ respectively, used to obtain the element following this equation.

For all these attitudes, whenever a face appears brighter, it proportionally appears smaller from the observer's point of view, and vice-versa to compensate, which gives intuition to the fact that swapping the $\hat{v}_B$ and $\hat{s}_B$ vectors results in the same spectrum observation.


\subsection{Phase dispersion minimisation}

The results in Table \ref{tab:PDM} show, for each dataset, the fraction of the times when the $k$-th PDM estimate of the period was the first to be within 5\% of the real value. They show that for dataset 2, there is a high failure rate for this overly simplistic method, and that when the lowest optimum does not result in the real estimate, subsequent optima are not accurate either. 
More sophisticated methods exist in the literature, such as generalised Lomb-Scargle \cite{zechmeister2009generalised}.
It the future, it may be useful to incorporate this or other methods into our approach.
However, as shown in subsequent sections, at the moment the best first DDP method produces good results without requiring knowledge of the period beforehand, so this was not included in this phase of the work.

\begin{table}[thb]
\centering
\caption{For both datasets, the fraction of times the $k$-th local minimum of the sum of variances of same phase samples is within 5\% of the correct value}
\label{tab:PDM}
\begin{tabular}{@{}lrrrr@{}}
\toprule
& Dataset 1 & \multicolumn{3}{c}{Dataset 2} \\
$k$ &   & $J^{(1)}$ & $J^{(2)}$ & $J^{(3)}$ \\
\midrule
$1$ & $0.95$  & $0.65$ & $0.68$ & $0.67$ \\
$2\leq k \leq 5$ & $0.00$  & $0.00$ & $0.00$ & $0.00$ \\
$\geq 6$ & $0.05$ & $0.35$ & $0.32$ & $0.33$\\
\bottomrule
\end{tabular}
\end{table}

\subsection{Regularised Least Squares}

To separate the effect of the accuracy PDM, the procedure for obtaining the initial guess outlined in Section \ref{sec:lightcurve:initial_guess} was started from the correct rotation period, as opposed to using the results of PDM.
The methods described in Section \ref{sec:lightcurve:reg_LS} were then used to optimise the attitude history. The loss function being minimised is 
\begin{equation} \label{eq:loss}
    L({\bf q}(t)) = 
    \sum_i
    E(S({\bf v}_b(t_i), {\bf s}_b(t_i)), \hat{S}_i)
    +
    \eta
    J_\alpha
    ~,
\end{equation}
where $J_\alpha$, given by Eq. \ref{eq:J_alpha}, is the cost function penalising angular acceleration.
The value of $\eta$ used was $1 ~{\rm rad}^{-1}~s^2$ in all experiments.

For dataset 1, the methodology was applied to tracklets $\hat{S}^{(k)}_i$ for $k={1,\cdots,6}$, which correspond to 6 tracklets obtained at different times of year, chosen as the first observation tracklet obtained by simulation for the months of February, March, April, September, October, and November of 2023. For the other months of the same year, the observation conditions did not allow a single tracklet, in the simulations.
For dataset 2, the tracklets $\hat{S}^{(k)}_i$ correspond to random different orbits and angular velocities all obtained for the month of January of 2023.

The accuracy is measured as the root mean squared error of the angle $\theta$ of the rotation that goes from the ground truth ${\rm q}_{GT}(t_i)$ attitude to the estimated value ${\bf q}(t_i)$, $RMS(\theta)$.



For dataset 1, the grid search in Section \ref{sec:lightcurve:initial_guess} was applied to obtain an initial guess. Let ${\bf q}_0(t)$ be the initial guess that produces the lowest cost. 
The elements of the symmetry set defined with respect to this initial guess, $\mc{R}({\bf q}_0(t))$ in Eq. (\ref{eq:symm_group}), often converge to a better solution than if one started only from ${\bf q}_0(t)$.
Therefore, all of the elements of $\mc{R}({\bf q}_0(t))$ are refined by optimising the regularised least squares loss $L$ in Eq. (\ref{eq:loss}) with the constraint of fixed axis rotation, and applied only to a segment of length $T$, where $T$ is the estimated period of rotation.
Whichever produces the lowest cost is then used as an initial guess to optimise the same loss function $L$ without a fixed axis constraint.
Afterwards, both the interior point algorithm and DDP were used to optimise starting from this initial guess. 
The best first search is not being used at this point in the results.
Both methods were found to converge to the exact solution each time, although in most cases, there would not have been convergence if we were not using all elements of $\mc{R}({\bf q}_0(t))$ as initial guess for optimisation, where ${\bf q}_0(t)$ is the initial guess defined in Section \ref{sec:lightcurve:initial_guess}. 
Only two of the lightcurves in dataset 1, and one of them in dataset 1A converge when starting from ${\bf q}_0(t)$. 
All others only converged when the search was started from a different element of $\mc{R}({\bf q}_0(t))$.



Because the attitude histories in set $\mc{R}$ of Eq. (\ref{eq:symm_group}) all result in the same spectrum, and thus are indistinguishable, one should interpret the solution of a method not as a single attitude history $\bf q$, but as the set $\mc{R}({\bf q})$, which, if the method converged properly, should contain the ground truth ${\bf q}_{GT}(t)$.
Therefore, comparisons with ground truth should be done for all elements of $\mc{R}({\bf q}_{GT}(t))$, so as to judge the performance of a method correctly.

Despite this, for test case 1 and 1A, without noise in the data, of the attitude histories in $\mc{R}({\bf q}_{GT}(t))$, only the ground truth ${\bf q}_{GT}(t)$ is a global optimum.
The same does not occur with dataset 2.
This is because if ${\bf q}_{GT}(t)$ corresponds to fixed axis rotation, the value of $J_\alpha$ will be zero for this attitude history, while in general the other elements of $\mc{R}({\bf q}_{GT}(t))$ will not, and so the $J_\alpha$ loss will be higher.
For test case 1, our method always estimated an attitude history nearly identical to the ground truth, however, for 1A, one of the six tests performed converged to one of the attitude histories in $\mc{R}({\bf q}_{GT}(t))/\{{\bf q}_{GT}(t)\}$, as shown in Table \ref{tab:RMSth}, where the elements of $\mc{R}({\bf q}_{GT}(t))$ 
by the different elements $T_H$ and $T_B$, of $\mc{T}_H$ and $\mc{T}_B$ respectively, used in Eq. (\ref{eq:symm_group}) to specify an element $\mc{R}$.
Therefore, this method tends to converge to a fixed axis rotation solution if one exists that explains the observed spectra.

\begin{figure}[htb]
    \centering    
    \includegraphics[width=0.9\textwidth]{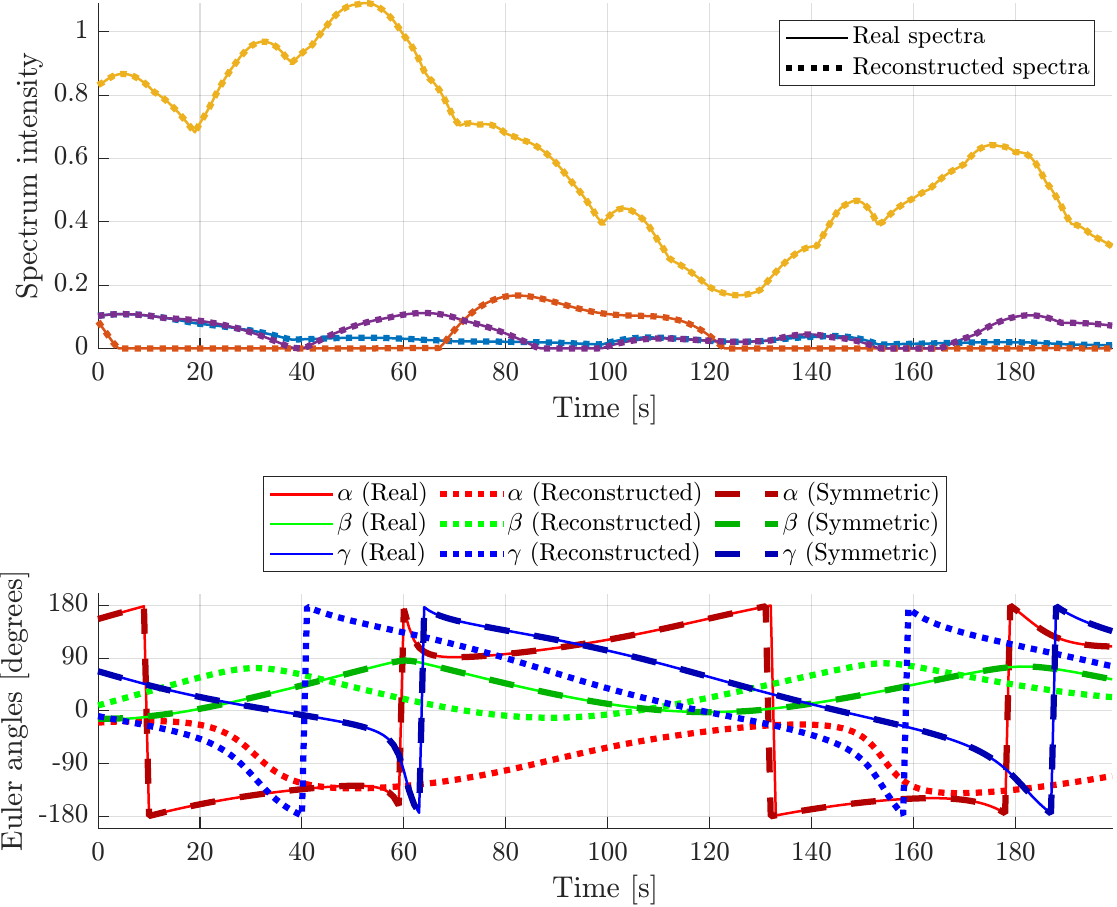}
    \caption{Top: Normalised spectrum intensity and reconstructed attitude time series for dataset 2, with $\hat{S}^{(2)}$ and $J^{(2)}$. Each colour represents a different spectral component. The plots overlap almost exactly. 
    Bottom: Angles $\alpha$, $\beta$, and $\gamma$ are ZYX Euler angles, obtained using Matlab's function \textit{quat2eul}.}
    \label{fig:spectr_recon_symm}
\end{figure}

For dataset 2, the higher difficulty associated with this test case meant that neither of the algorithms were converging as the initial guesses were often too far from the ground truth.
Therefore, for dataset 2, only the results using the best-first DDP method of Section \ref{sec:lightcurve:DDP} are shown, in Table \ref{tab:DDP_DSet2_J_all}, which shows the $RMSE(\theta)$ between each of the predicted attitudes and the closest element of $\mc{R}({\bf q}_{GT}(t))$, alongside the symmetry transformations associated with that closest element.
Figure \ref{fig:spectr_recon_symm} shows the case with $\hat{S}^{(2)}$ and $J^{(2)}$, which converged to a solution that is different from ground truth, as shown in Table \ref{tab:DDP_DSet2_J_all}.
The real spectra matched the spectra obtained from the estimated attitude time series almost exactly, but the attitude time series itself, represented with $XYZ$ Euler angles, does not match the ground truth.
The symmetric version, however, matches very closely.

\begin{table}[thb]
\centering
\caption{ Values of $RMS(\theta)$ for test case 1A comparing estimated attitude to various elements of $\mc{R}({\bf q}_{GT}(t))$}
\label{tab:RMSth}
\begin{tabular}{@{}lrrrr@{}}
\toprule
$T_b$ & \multicolumn{2}{c}{$I$} & \multicolumn{2}{c}{$T_{BR}$} \\ 
$T_h$ &     $I$ & $T_{H_3}$ & $T_{H_2}$ & $T_{H_1}$\\ \midrule
$\hat{S}^{(1)}$ & $\bf 0.00^\circ$ & $180.00^\circ$ & $91.43^\circ$ & $114.38^\circ$ \\
$\hat{S}^{(2)}$ & $\bf 0.00^\circ$ & $180.00^\circ$ & $111.12^\circ$ & $110.83^\circ$ \\
$\hat{S}^{(3)}$ & $122.09^\circ$ & $126.39^\circ$ & $\bf 1.47^\circ$ & $179.52^\circ$ \\
$\hat{S}^{(4)}$ & $\bf 0.00^\circ$ & $180.00^\circ$ & $125.67^\circ$ & $122.00^\circ$ \\
$\hat{S}^{(5)}$ & $\bf 0.00^\circ$ & $180.00^\circ$ & $112.13^\circ$ & $113.19^\circ$ \\
$\hat{S}^{(6)}$ & $\bf 0.00^\circ$ & $180.00^\circ$ & $76.93^\circ$ & $135.52^\circ$ \\\bottomrule
\end{tabular}
\end{table}

\begin{table}[thb]
\centering
\caption{For dataset 2, the value of $RMS(\theta)$ is reported alongside the symmetry transformations $T_B$ and $T_H$ that correspond to the element of $\mc{R}({\bf q}_{GT}(t))$ that was closest to the estimated attitude time series.}
\label{tab:DDP_DSet2_J_all}
\begin{tabular}{@{}lccccccccc@{}}
\toprule
 & \multicolumn{3}{c}{$J^{(1)}$} & \multicolumn{3}{c}{$J^{(2)}$}  & \multicolumn{3}{c}{$J^{(3)}$} \\
  & $T_B$ & $T_H$ & $RMS(\theta)$ & $T_B$ & $T_H$ & $RMS(\theta)$ & $T_B$ & $T_H$ & $RMS(\theta)$ \\
\midrule
$\hat{S}^{(1)}$ & $I$ & $I$ & $0.00^\circ$        & $I$      & $I$       & $0.03^\circ$ &  $I$      & $T_{H_3}$ & $5.04^\circ$ \\
$\hat{S}^{(2)}$ & $I$ & $I$ & $0.02^\circ$        & $I$      & $T_{H_3}$ & $0.38^\circ$ &  $I$      & $I$       & $0.12^\circ$ \\
$\hat{S}^{(2)}$ & $I$ & $I$ & $0.00^\circ$        & $I$      & $I$       & $0.04^\circ$ &  $I$      & $I$       & $0.03^\circ$ \\
$\hat{S}^{(4)}$ & $I$ & $I$ & $0.09^\circ$        & $I$      & $T_{H_3}$ & $0.22^\circ$ &  $I$      & $T_{H_3}$ & $0.10^\circ$ \\
$\hat{S}^{(5)}$ & $I$ & $I$ & $0.02^\circ$        & $I$      & $I$       & $0.14^\circ$ &  $I$      & $I$       & $0.06^\circ$ \\
$\hat{S}^{(6)}$ & $I$ & $I$ & $0.87^\circ$        & $I$      & $I$       & $1.89^\circ$ &  $I$      & $I$       & $0.88^\circ$ \\
\bottomrule
\end{tabular}
\end{table}

\subsection{Machine Learning Results}
\label{sec:results:ML}

To determine the best choice of reference frame for the variables and the best choice of network architecture,
the dataset 1 is used to train and validate the various ML models under consideration.
Because the datasets are of varying lengths, while the network accepts fixed length data, some pre-processing is required.
The length used is 200 seconds, with one sample per second.

The data is split chronologically, with the training data being composed of the first 70\% of the data, and the validation data is the remaining last 30\%.
The chronological split serves to verify the model's ability to generalise from the available data. 




\subsubsection{Choice of Reference Frame}

For this test, the results with different reference frames for the input vectors $\hat{\bf v}$ and $\hat{\bf s}$ are compared in the results of predicting both $\boldsymbol{\omega}_{IB}$ and $\boldsymbol{\omega}_{BI}$.

\begin{figure}[thb]
    \centering
    \includegraphics[width=0.8\textwidth]{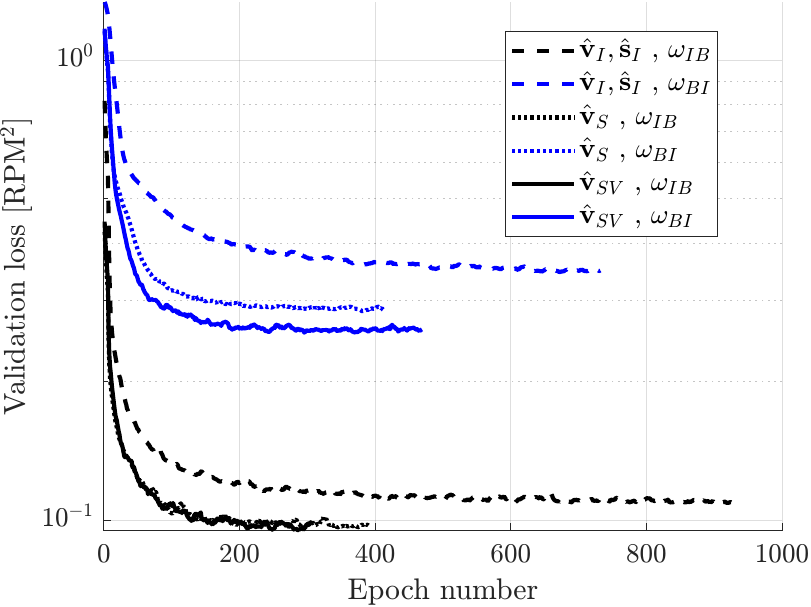}
    \caption{Validation loss as a function of epoch number. 
    Tests for different reference frames.
    To make the plots visually easier to interpret, a moving average filter with a window size of 10 was used.
    In the legend, the inputs and outputs of the model are used to identify the plots.}
    \label{fig:ML_Test_1}
\end{figure}

Figure \ref{fig:ML_Test_1} shows the validation losses as a function of epoch.
To make the plots visually easier to interpret, a moving average filter with a window size of 10 was used.
In the legend, the inputs and outputs of the model are used to identify the plots.
From the results of this test, in Fig. \ref{fig:ML_Test_1}, $\boldsymbol{\omega}_{IB}$ clearly seems easier to predict than $\boldsymbol{\omega}_{BI}$, but it is necessary to test whether this is due to the symmetries which, for the object in this test, are only present in the inertial frame.

\subsubsection{Handling Symmetry}
\label{sec:results:ML:symm}

The fact that, for an asymmetric object, the symmetries only occur in the I frame may seem like it is best to simply predict $\boldsymbol{\omega}_{IB}$ instead, but these symmetries are independent of the object, and thus easier to handle in general. 
Furthermore, if the attitude of the object is not available, $\boldsymbol{\omega}_{IB}$ and $\boldsymbol{\omega}_{BI}$ convey different information, the former is the angular velocity in a body frame, while the latter is in the inertial frame.

\begin{figure}[thb]
    \centering
    \includegraphics[width=0.8\textwidth]{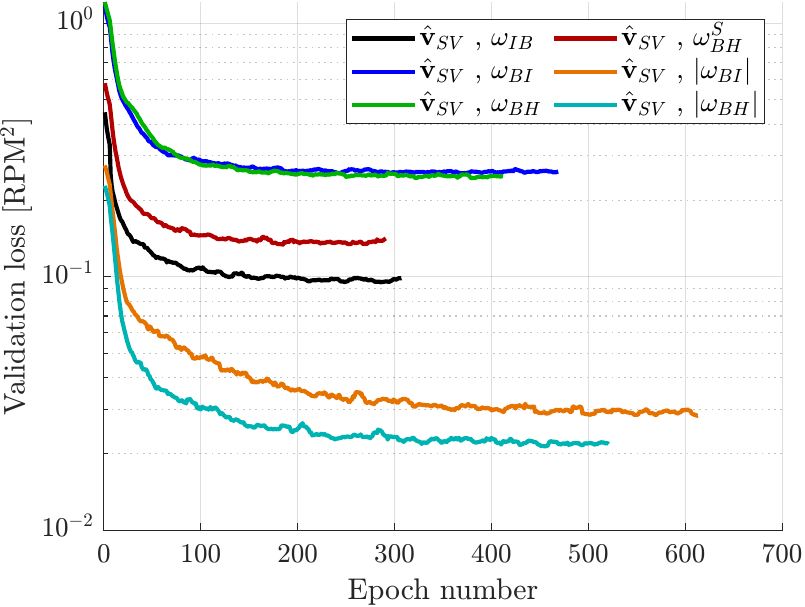}
    \caption{Validation loss as a function of epoch number. 
    Tests for different output forms, testing ways to handle symmetry in data.
    }
    \label{fig:ML_Test_2}
\end{figure}

In Section \ref{sec:lightcurve:symm}, the idea of using the H frame to handle symmetries was proposed. Plots of the validation loss when the model is trained on $\boldsymbol{\omega}_{BI}$, $\boldsymbol{\omega}_{IB}$, and $\boldsymbol{\omega}_{BH}$ are compared in Fig. \ref{fig:ML_Test_2}. In the same figure, $\boldsymbol{\omega}_{BH}^S$ denotes $\boldsymbol{\omega}_{BH}$ with its $y$ and $z$ components replaced by their absolute values, while $\lvert \boldsymbol{\omega}_{BI} \rvert$ and $\lvert \boldsymbol{\omega}_{BH} \rvert$ denote vectors with all their components replaced by their absolute values.

A less expected result is that $\lvert \bs{\omega}_{BH} \rvert$ performs much better than $\bs{\omega}_{BH}^S$, where the only difference between them is that in $\bs{\omega}_{BH}^S$, the $x$ component is not the absolute value and may be negative.
This cannot be explained by the theoretical discussion in Section \ref{sec:lightcurve:symm}, as the symmetries should not affect the ability of the method to accurately measure the sign of the $x$ component of $\bs{\omega}_{BH}$.
On the other hand, the results with numerical optimisation methods always show a convergence to one of the identified elements of the set $\mc{R}({\bf q}_{GT}(t))$, which would suggest a symmetric case was not missed.
Therefore, the cause for the much better performance of predicting $\lvert \bs{\omega}_{BH} \rvert$ over $\bs{\omega}_{BH}^S$ with ML methods warrants further study. 

Nevertheless, the fact that using $\lvert \bs{\omega}_{BH} \rvert$ is significantly better than using $\lvert \bs{\omega}_{BI} \rvert$ further confirms that explicitly handling the ambiguity of the lightcurve inversion process caused by the presence of symmetries does improve the results.

\begin{figure}
    \centering
    \begin{subfigure}[b]{0.49\textwidth}
        \centering
        \includegraphics[trim={50 0 50 0}, clip, width=\textwidth]{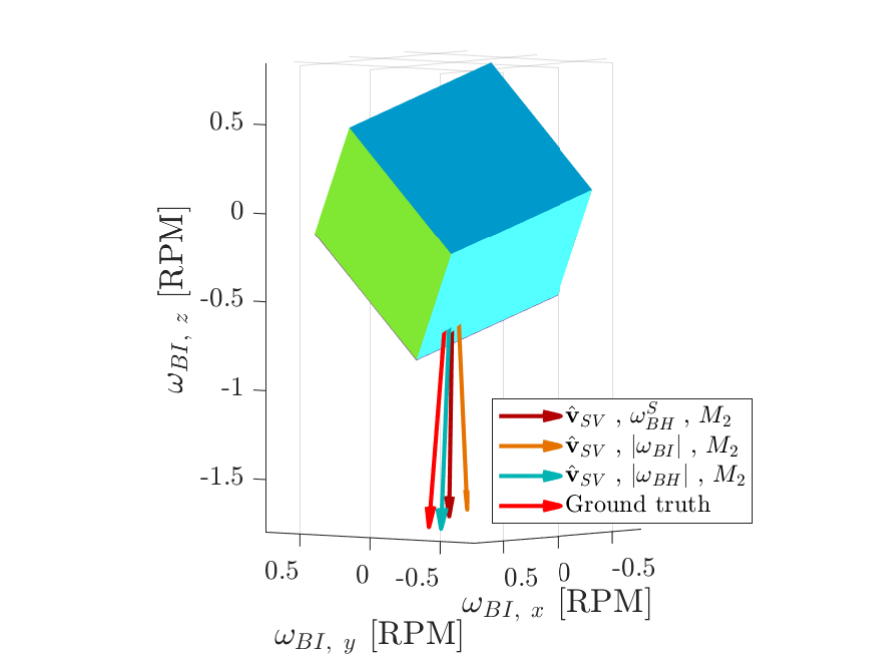}    
        \caption{Dataset 1 using $M_2$ architecture}
    \end{subfigure}
    \hfill
    \begin{subfigure}[b]{0.49\textwidth}
    \centering
        \includegraphics[trim={50 0 50 0}, clip, width=\textwidth]{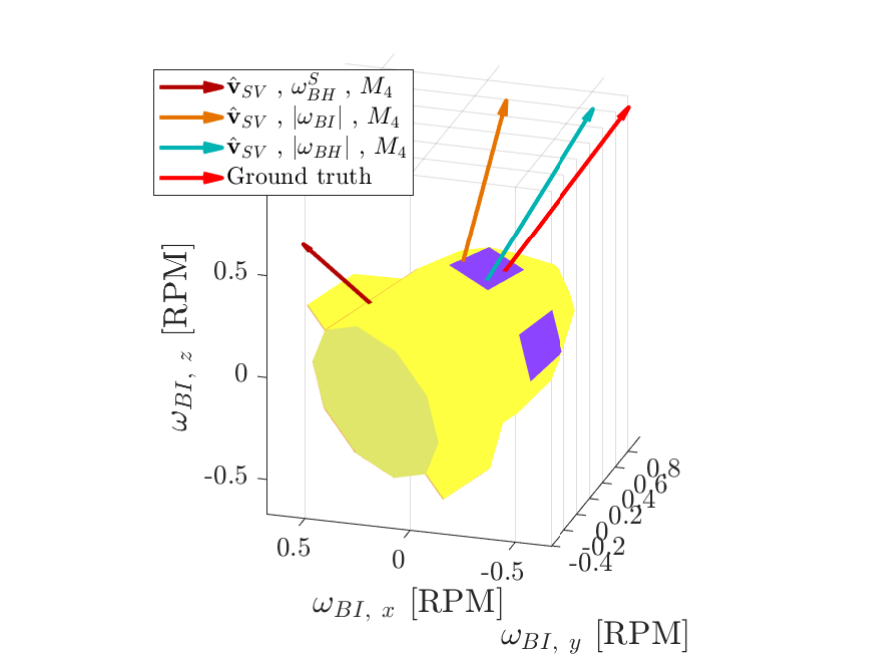}    
        \caption{Dataset 2 using $M_4$ architecture}
    \end{subfigure}
    \caption{Angular velocity vectors, converted to inertial (ECI) frame, as predicted by neural networks trained in different configurations}
    \label{fig:vectors}
\end{figure}

Apart from $\bs{\omega}_{IB}$, all other outputs of the model can be used to obtain one or more candidate values for $\bs{\omega}_{BI}$. 
When the output is $\bs{\omega}_{BH}$, for instance, Eq. (\ref{eq:wBH}) can be solved to obtain $\bs{\omega}_{BI}$.
If either $\bs{\omega}_{BH}^S$, $\lvert\bs{\omega}_{BI}\rvert$ or $\lvert\bs{\omega}_{BH}\rvert$ are being used, the elements whose absolute value was taken may have either sign in reality.
Figure \ref{fig:vectors} shows two examples of the predicted $\bs{\omega}_{BI}$, one for each dataset.
For the ambiguous predictions, the combination of signs that results in $\bs{\omega}_{BI}$ closest to ground truth is shown.
The legend represents what was the input and output of the model, followed by its architecture, like in other figures. Nonetheless, what is being shown is always the vector $\bs{\omega}_{BI}$ that corresponds to each specific prediction.

\subsubsection{Choice of Network Architecture}

\begin{figure}
    \centering
    \includegraphics[width=0.9\textwidth]{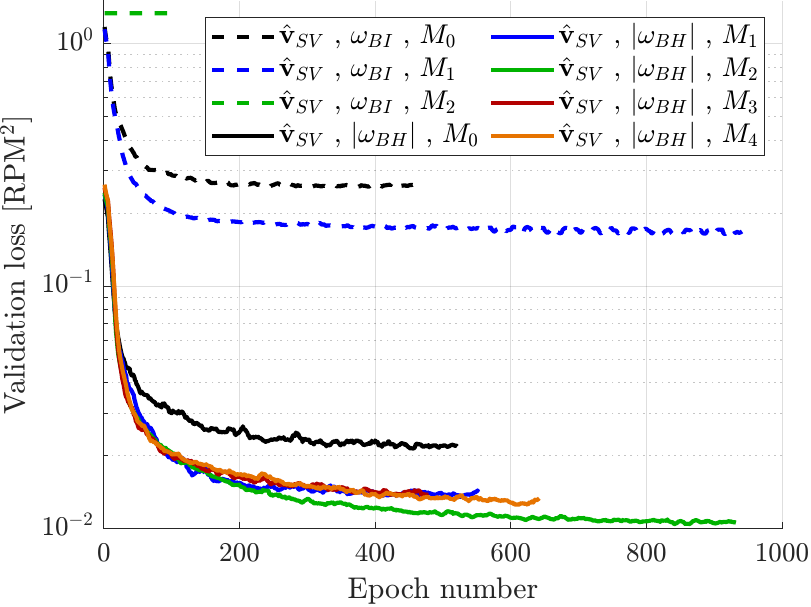}
    \caption{Validation loss as a function of epoch number. 
    Tests for different network architectures with dataset 1.
    }
    \label{fig:ML_Test_3}
\end{figure}

\begin{figure}[thb]
    \centering
    \includegraphics[width=0.8\textwidth]{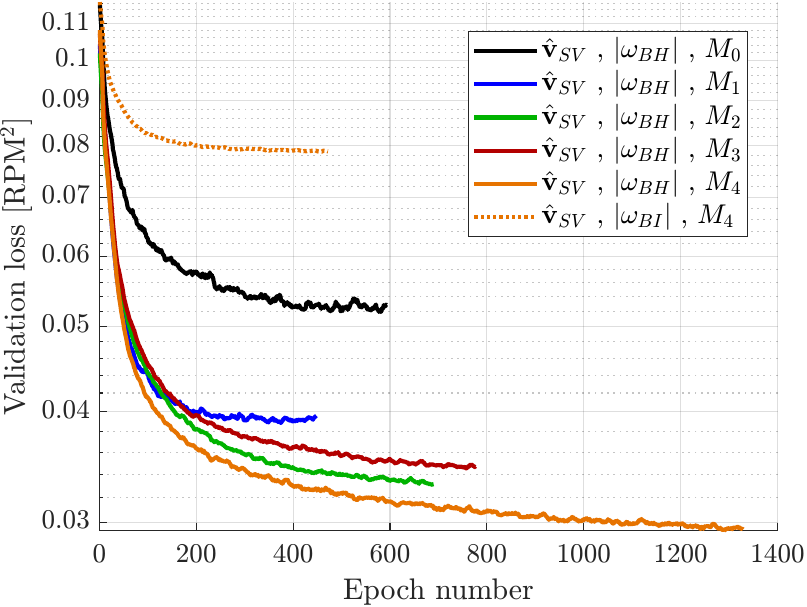}
    \caption{Validation loss as a function of epoch number. 
    Tests for different network architectures with dataset 2 and predicting $\lvert \bs{\omega}_{BH} \rvert$. For comparison, results when predicting $\lvert \bs{\omega}_{BI} \rvert$ are shown only for the network architecture with best results, $M_4$.
    }
    \label{fig:ML_Test_4}
\end{figure}

The network architectures defined in Section \ref{sec:lightcurve:ML} are now compared.
Tests above were only for dataset 1. Here, results for dataset 1 are in Fig. \ref{fig:ML_Test_3} and for dataset 2 in Fig. \ref{fig:ML_Test_4}.
For dataset 1, a much simpler test case, adding more layers or neurons after model $M_2$ does not bring additional advantage, while for dataset 2 additional model complexity did translate into improved results, a trend that may have continued had there been time to train even more complex models on this data.

\subsection{Test in Laboratory Environment}
\label{sec:results:ML:lab}

As a first step towards using these attitude determination strategies with real data, lab observations obtained for a rotating painted cube was used.
The painting cube data was used because this allows using a Lambertian model, which is easier to work with than specular models.

\begin{figure}[htb]
     \centering
     \begin{subfigure}[b]{0.47\textwidth}
         \centering
         \includegraphics[width=\textwidth]{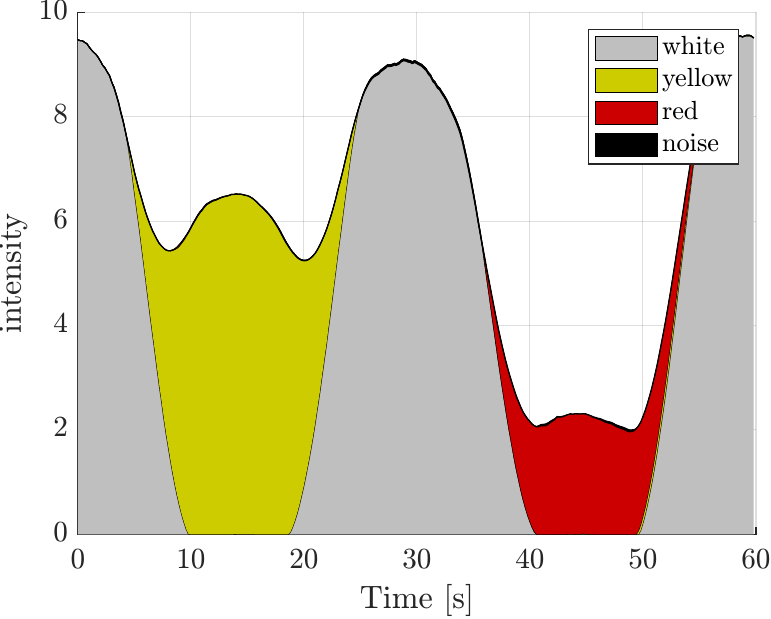}
         \caption{Cube with faces painted in sequence white-yellow-red-white}
         \label{fig:lab:WYR}
     \end{subfigure}
     \hfill
     \begin{subfigure}[b]{0.47\textwidth}
         \centering
         \includegraphics[width=\textwidth]{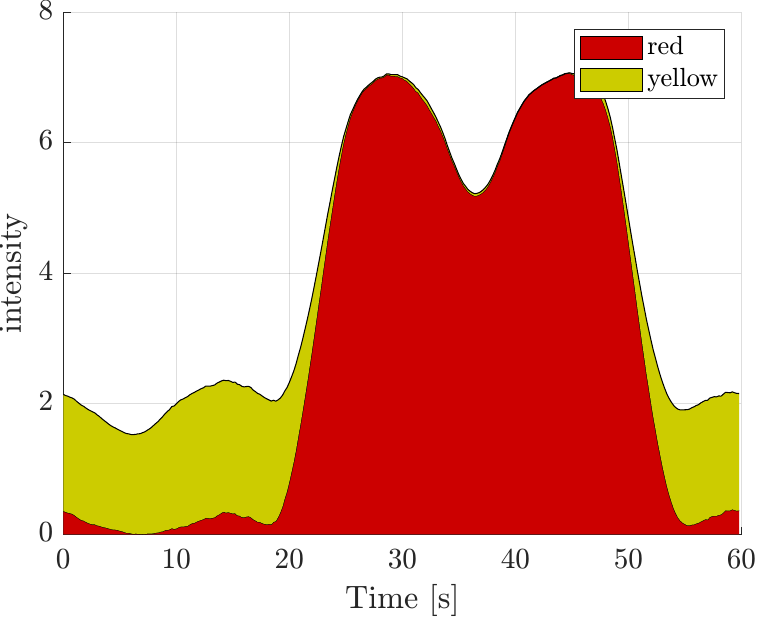}
         \caption{Cube with faces painted in sequence yellow-yellow-red-red}
         \label{fig:lab:RY}
     \end{subfigure}
     \caption{Unmixed spectra for lab data of painted cubes}
     \label{fig:lab}
\end{figure}

Data pertaining to two different painted cubes was used.
Only four faces of this cube are imaged, as the cube rotates about its Z axis, and the observation and illumination are done at an axis perpendicular to it. The phase angle is 10 degrees.
The first cube has the faces coloured in order white-yellow-red-white, while the second cube has yellow-yellow-red-red.
The spectral data, unmixed into the material components (white, yellow, and red paints), are shown in Fig \ref{fig:lab}. 
The noise component in these figures represents the residual after the unmixing. This noise component is very small compared to the measured signal.

\begin{figure}[htb]
     \centering
     \begin{subfigure}[b]{0.47\textwidth}
         \centering
         \includegraphics[width=\textwidth]{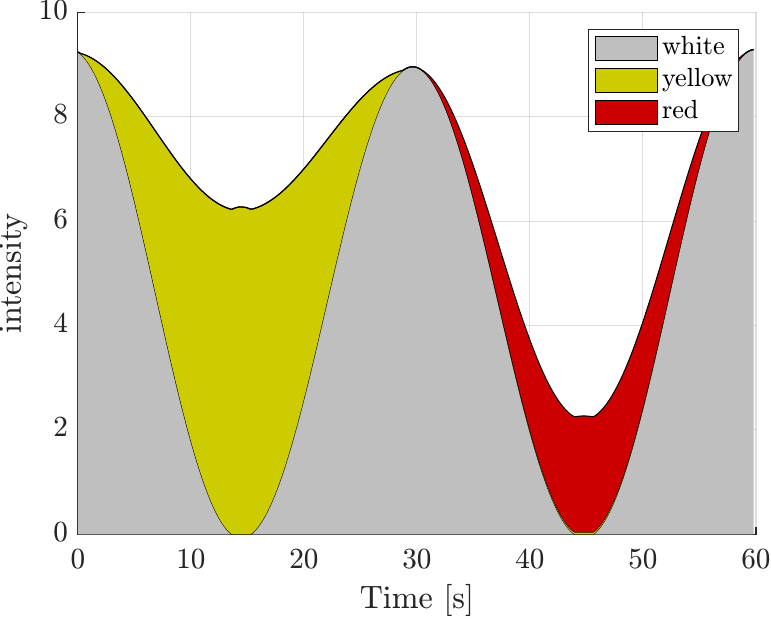}
         \caption{Cube with faces painted in sequence white-yellow-red-white}
         \label{fig:lab_A:WYR}
     \end{subfigure}
     \hfill
     \begin{subfigure}[b]{0.47\textwidth}
         \centering
         \includegraphics[width=\textwidth]{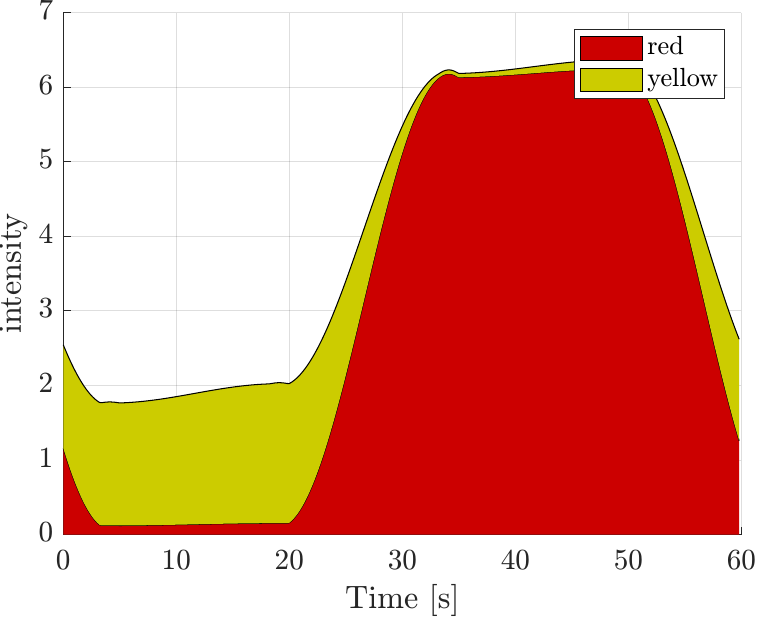}
         \caption{Cube with faces painted in sequence yellow-yellow-red-red}
         \label{fig:lab_A:RY}
     \end{subfigure}
     \caption{Reconstructed spectra following least squares approach with known fixed axis rotation}
     \label{fig:lab_A}
\end{figure}

As an initial experiment, the least squares methodology used to find an initial guess, i.e. using a fixed axis constraint, is used. 
The parameters being determined with this method are the angular velocity and the initial angle, in addition to constant coefficients for the intensity of each face.
Figure \ref{fig:lab_A} shows the spectra reconstructed from the observations.

The curves of these figures look substantially different from the lab data.
This is because the spectral model in Eq. \ref{eq:Lambert} is not adequate to model these observations, due to the fact that the field of view (FOV) of the camera does not contain the entire object.
This was an intentional choice, which contributed to the low-noise present in the observations.

To obtain a better fit, the spectral simulation function was adapted to model this effect.
The formula is obtained by replacing, in Eq. (\ref{eq:Lambert}), $-\hat{{\bf v}}\cdot\hat{\mathbf{n}}_i$ by $F({\bf v}, \hat{\mathbf{n}}_i, V_{ij}, \nu, D)$, 
where $V_{ij}$ are the coordinates of the $j$-th vertex of the $i$-th face, $\nu$ is the field of view angle, and $D$ is the distance to the object.

The function $F({\bf v}, \hat{\mathbf{n}}_i, V_{ij}, \nu, D)$ obtains the apparent area of the intersection of the $i$-th face with the field of view cone. 
This is done by projecting the points in $V_{ij}$ into the image plane, the plane orthogonal to ${\bf v}$, and finding the area, in this plane, of its intersection with the projection of the FOV onto the same plane, a circle of radius $\nu$. This area is the value of $F({\bf v}, \hat{\mathbf{n}}_i, V_{ij}, \nu, D)$. 
Note that if the FOV fully contains the face, $F({\bf v}, \hat{\mathbf{n}}_i, V_{ij}, \nu, D)$ will be approximately proportional to $-\hat{{\bf s}}\cdot\hat{\mathbf{n}}_i$ for large distances $D$.
\footnote{The code for this approach was developed based on this Stack Overflow answer: https://stackoverflow.com/a/22420313/1577940}

\begin{figure}[htb]
     \centering
     \begin{subfigure}[b]{0.47\textwidth}
         \centering
         \includegraphics[width=\textwidth]{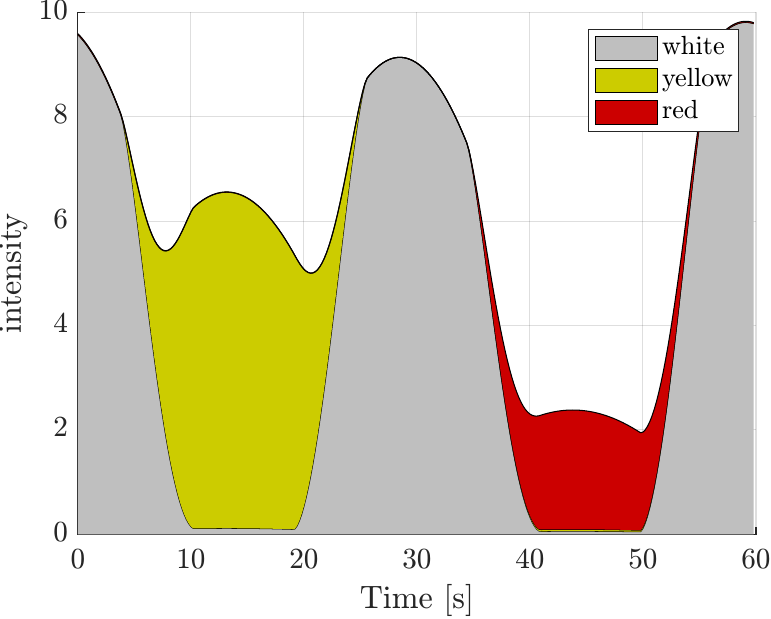}
         \caption{Cube with faces painted in sequence white-yellow-red-white}
         \label{fig:lab_B:WYR}
     \end{subfigure}
     \hfill
     \begin{subfigure}[b]{0.47\textwidth}
         \centering
         \includegraphics[width=\textwidth]{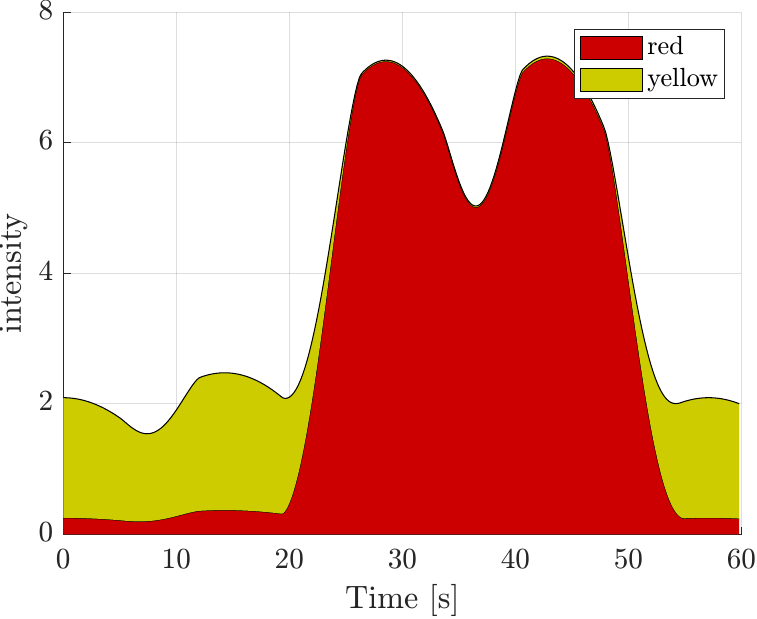}
         \caption{Cube with faces painted in sequence yellow-yellow-red-red}
         \label{fig:lab_B:RY}
     \end{subfigure}
     \caption{Reconstructed spectra following least squares approach with known fixed axis rotation, using FOV model.}
     \label{fig:lab_B}
\end{figure}

Using this model, the reconstructed spectrum from the reconstructed attitude history is shown in Fig. \ref{fig:lab_B}, which now resembles much more closely the lab data in Fig. \ref{fig:lab}.

The full regularised least squares method of Section \ref{sec:lightcurve:reg_LS}, i.e. without assuming known fixed axis of rotation, was not used, as this model is much more computationally expensive, and those approaches require a large number of computations of $S$ in order to converge.
However, the new best-first search DDP method has not yet been tested on this data. 
\section{Conclusions}
\label{sec:conclusions}

Two approaches to estimate the rotation and/or attitude of a spacecraft from spectral lightcurve observations were proposed and tested.
One based on modelling the spectral observations and using regularised least squares to find the attitude history that matches it, and another used neural networks to estimate the angular velocity vector.

In both cases it was found that taking into account symmetries in the lightcurve function, which are a consequence of the scattering model used, is essential to obtain an accurate method, and also to test it correctly.
The tests with neural networks also showed that conditioning the data prior to training, in the form of reference frame choices, influences strongly the quality of the results one obtains.
However, there was an unexpected result with the ML approach, as taking the absolute value of all components of the angular velocity vector in the H frame produced significantly better results than only taking the absolute values of those components whose signs may change due to the symmetries.

This study of symmetries also shows that, without knowledge of the exact attitude dynamics the object is subject to, one cannot in general find a single solution to the attitude determination problem from spectral lightcurves, despite the introduction of time series data.

\bibliography{references}

\end{document}